\documentclass[twocolumn,amsmath,showpacs,amsfonts,aps,prc,floatfix]{revtex4}
\usepackage{graphicx}
\usepackage{bm}

\begin{document}

\title{Fluctuating initial conditions and fluctuations in elliptic and triangular flow}

\author{A. K. Chaudhuri}
\email[E-mail:]{akc@veccal.ernet.in}
\affiliation{Variable Energy Cyclotron Centre, 1/AF, Bidhan Nagar, 
Kolkata 700~064, India}

\begin{abstract}

In heavy ion collisions, event-by-event fluctuations in participating nucleon
positions can lead to triangular flow. With fluctuating  initial conditions, flow coefficients will also fluctuate.  
In a hydrodynamic model, we study the fluctuations in elliptic and triangular flow,  due to fluctuating initial conditions. Both elliptic and triangular flow fluctuates strongly, triangular flow more strongly than the elliptic flow. Strong fluctuations greatly reduce the sensitivity of elliptic and triangular flow to viscosity. 

 \end{abstract}


\date{\today}  

\maketitle


\section{Introduction} \label{intro}
 
 It is expected that collisions between two nuclei at ultra-relativistic energies will lead to a phase transition from hadrons to the fundamental constituents, quarks and gluons, usually referred to as Quark-Gluon-Plasma (QGP).  One of the experimental observables of QGP is the azimuthal distribution of produced particles.  In non-zero impact parameter collision between two identical nuclei, the collision zone is asymmetric. Multiple collisions transform the initial asymmetry   into momentum anisotropy. Momentum anisotropy is best studied by decomposing it   in a
   Fourier series,

\begin{equation}
\frac{dN}{d\phi}=\frac{N}{2\pi}[1+ 2\sum_n v_n cos(n\phi-n\psi_n)], n=1,2,3...
\end{equation} 
 
\noindent   $\phi$ is the azimuthal angle of the detected particle and 
$\psi_n$ is the  plane of the symmetry of initial collision zone. For smooth initial matter distribution, plane of symmetry of the collision zone coincides with the reaction plane (the plane containing the impact parameter and the beam axis), 
$\psi_n \equiv \Psi_{RP}, \forall n$. The odd Fourier coefficients are zero by symmetry. However, fluctuations in the positions of the participating nucleons can lead to non-smooth density distribution, which will fluctuate on event-by-event basis.  
The participating nucleons then determine the symmetry plane ($\psi_{PP}$), which fluctuate around the reaction plane \cite{Manly:2005zy}. As a result odd harmonics, which were exactly zero for smoothed initial distribution, can be developed. It has been conjectured that third harmonic $v_3$, which is response of the initial triangularity of the medium, is responsible for the observed structures in two particle correlation in Au+Au collisions \cite{Mishra:2008dm},\cite{Mishra:2007tw},\cite{Takahashi:2009na},\cite{Alver:2010gr},\cite{Alver:2010dn},\cite{Teaney:2010vd}. The ridge structure in pp collisions also has a natural explanation if odd harmonic flow develops.  Recently, ALICE collaboration has observed odd harmonic flows    in Pb+Pb collisions \cite{:2011vk}. In most central collisions, the elliptic flow ($v_2$) and triangular flow ($v_3$) are of similar magnitude. In peripheral collisions however, elliptic flow dominates.

The second harmonic  or the elliptic flow ($v_2$) has been studied extensively in $\sqrt{s}_{NN}$=200 GeV Au+Au collisions at RHIC \cite{PHENIXwhitepaper,STARwhitepaper}. Recently, ALICE collaboration measured elliptic flow in $\sqrt{s}_{NN}$=2.76 TeV Pb+Pb collisions at LHC \cite{Aamodt:2010pa,:2011vk}. Large elliptic flow  has provided compelling evidence that at RHIC and LHC, nearly perfect fluid is produced. Deviation from the ideal fluid behavior is controlled by shear viscosity to entropy ratio ($\eta/s$). Effect of shear viscosity is to dampen the flow coefficients. Elliptic flow   has sensitive dependence on  $\eta/s$. In smooth hydrodynamics,
sensitivity of elliptic flow has been utilised to obtain phenomenological estimates of $\eta/s$ \cite{Luzum:2008cw,Song:2008hj,Chaudhuri:2009uk,Chaudhuri:2009hj,Roy:2011xt,Schenke:2011tv,Bozek:2011wa,Song:2011qa}. Triangular flow is supposed to be more sensitive to viscosity than the elliptic flow \cite{Alver:2010dn,Teaney:2010vd} and one expects  that triangular flow measurements will constrain   $\eta/s$ more accurately.

Event-by-event fluctuations in initial conditions generate the triangular flow.
It is then natural that the triangular flow itself will also fluctuate, event-by-event. Unless the fluctuations are within some reasonable limit, sensitivity of the flow to $\eta/s$ will reduce greatly. 
In the present paper, in a hydrodynamic model,  we have studied the fluctuations in elliptic and triangular  flow due to fluctuating initial conditions. Its sensitivity to the $\eta/s$ is also studied. It appear that with fluctuating initial conditions, the sensitivity of elliptic and triangular flow to $\eta/s$ is greatly reduced. Triangular flow fluctuates more strongly than the elliptic flow and become even less sensitive to $\eta/s$ than the elliptic flow. Large fluctuations belie the possibility of constraining viscosity to entropy ratio from triangular flow measurements.

\section{Hydrodynamic equations, equation of state and initial conditions} \label{sec2}


We assume that in $\sqrt{s}_{NN}$=2.76 TeV, Pb+Pb collisions  at LHC, a baryon free fluid is formed.
Only dissipative effect we consider is the shear viscosity. Heat conduction and bulk viscosity is neglected.  
The space-time evolution of the fluid is obtained by solving,

\begin{eqnarray}  
\partial_\mu T^{\mu\nu} & = & 0,  \label{eq3} \\
D\pi^{\mu\nu} & = & -\frac{1}{\tau_\pi} (\pi^{\mu\nu}-2\eta \nabla^{<\mu} u^{\nu>}) \nonumber \\
&-&[u^\mu\pi^{\nu\lambda}+u^\nu\pi^{\mu\lambda}]Du_\lambda. \label{eq4}
\end{eqnarray}

Eq.\ref{eq3} is the conservation equation for the energy-momentum tensor, $T^{\mu\nu}=(\varepsilon+p)u^\mu u^\nu - pg^{\mu\nu}+\pi^{\mu\nu}$, 
$\varepsilon$, $p$ and $u$ being the energy density, pressure and fluid velocity respectively. $\pi^{\mu\nu}$ is the shear stress tensor. Eq.\ref{eq4} is the relaxation equation for the shear stress tensor $\pi^{\mu\nu}$.   
In Eq.\ref{eq4}, $D=u^\mu \partial_\mu$ is the convective time derivative, $\nabla^{<\mu} u^{\nu>}= \frac{1}{2}(\nabla^\mu u^\nu + \nabla^\nu u^\mu)-\frac{1}{3}  
(\partial . u) (g^{\mu\nu}-u^\mu u^\nu)$ is a symmetric traceless tensor. $\eta$ is the shear viscosity and $\tau_\pi$ is the relaxation time.  It may be mentioned that in a conformally symmetric fluid relaxation equation can contain additional terms  \cite{Song:2008si}. Assuming boost-invariance, Eqs.\ref{eq3} and \ref{eq4}  are solved in $(\tau=\sqrt{t^2-z^2},x,y,\eta_s=\frac{1}{2}\ln\frac{t+z}{t-z})$ coordinates, with the code 
  "`AZHYDRO-KOLKATA"', developed at the Cyclotron Centre, Kolkata.
 Details of the code can be found in \cite{Chaudhuri:2008sj}. 

\begin{figure}[t]
 \center
 \resizebox{0.30\textwidth}{!}{%
  \includegraphics{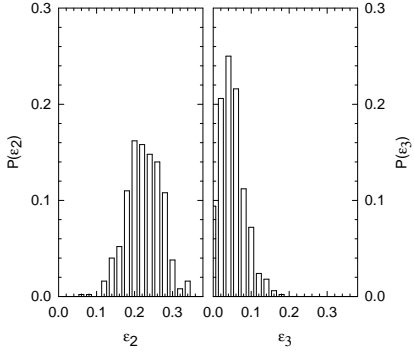}
}
\caption{Distribution of  spatial eccentricity ($\epsilon_2$) and triangularity ($\epsilon_3$) in b=8.9 fm Pb+Pb collisions in a model with hot spot formation. Hot spot width $\sigma$=1 fm.}
\label{F1}
\end{figure}

Eqs.\ref{eq3},\ref{eq4} are closed with an equation of state (EoS) $p=p(\varepsilon)$.
Currently, there is consensus that the confinement-deconfinement transition is a cross over and the cross over or the pseudo critical temperature for the  transition  is
$T_c\approx$170 MeV \cite{Aoki:2006we,Aoki:2009sc,Borsanyi:2010cj,Fodor:2010zz}.
In the present study, we use an equation of state where the Wuppertal-Budapest \cite{Aoki:2006we,Borsanyi:2010cj} 
lattice simulations for the deconfined phase is smoothly joined at $T=T_c=174$ MeV, with hadronic resonance gas EoS comprising all the resonances below mass $m_{res}$=2.5 GeV. Details of the EoS can be found in \cite{Roy:2011xt}.
  
Solution of partial differential equations (Eqs.\ref{eq3},\ref{eq4}) requires initial conditions, e.g.  transverse profile of the energy density ($\varepsilon(x,y)$), fluid velocity ($v_x(x,y),v_y(x,y)$) and shear stress tensor ($\pi^{\mu\nu}(x,y)$) at the initial time $\tau_i$. One also need to specify the viscosity ($\eta$) and the relaxation time ($\tau_\pi$). A freeze-out prescription is also needed to convert the information about fluid energy density and velocity to particle spectra.  We assume that the fluid is thermalised at $\tau_i$=0.6 fm and the initial fluid velocity is zero, $v_x(x,y)=v_y(x,y)=0$. We initialise the
shear stress tensor  to boost-invariant values, $\pi^{xx}=\pi^{yy}=2\eta/3\tau_i$, $\pi^{xy}$=0 and for  the relaxation time, we use the   Boltzmann estimate $\tau_\pi=3\eta/2p$. We also assume that the viscosity to entropy density ($\eta/s$) remains a constant throughout the evolution and simulate Pb+Pb collisions for a range of $\eta/s$. The freeze-out is fixed at $T_F$=130 MeV.

\begin{figure}[t]
 \center
 \resizebox{0.30\textwidth}{!}{%
  \includegraphics{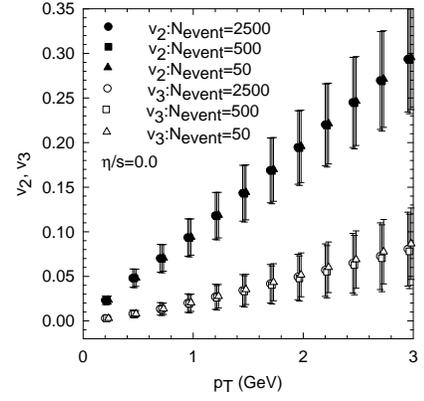} 
}
\caption{comparison of elliptic and triangular flow in hydrodynamical simulations with $N_{event}$=2500, 500 and 50. Data points are marginally shifted to distinguish between them.}
\label{F2}
\end{figure} 

For the fluctuating initial energy density distribution, we use a model of hot spots in the initial states.  
Recently, similar models are used to study   elliptic flow in pp collisions at LHC \cite{CasalderreySolana:2009uk,Bozek:2009dt,Chaudhuri:2009yp}. In \cite{Bhalerao:2011bp}, a similar model was used to study anisotropy in heavy ion collisions due to fluctuating initial conditions. 
We assume that in an impact parameter ${\bf b}$ collision, each participating nucleon pair randomly deposit some energy in the reaction volume, and produces a hot spot. The hot spots are assumed to be Gaussian distributed. The initial energy density is then super position of $N=N_{participant}$ hot spots. 
 
\begin{equation}
\varepsilon({x,y})=\varepsilon_0 
\sum_{i=1}^{N_{participant}} e^{-\frac{({\bf r}-{\bf r}_i)^2}{2\sigma^2}}
\end{equation}

The participant number $N_{participant}$ is calculated in a Glauber model. We also restrict the centre of hotspots ($r_i$) within the transverse area defined by the Glauber model of participant distribution. The central density $\varepsilon_0$ and the width $\sigma$ are parameters of the model. We fix $\sigma$=1 fm. The central density
$\varepsilon_0$ is fixed to reproduce approximately the experimental charged particles in a peripheral (30-40\%) Pb+Pb collisions. 

Since hot spot positions are random, density distribution     fluctuates from event to event.  Asymmetry in   the initial energy density distribution can be characterised in terms   of $\epsilon_n$ and $\psi_n$ \cite{Alver:2010gr} , 

\begin{equation} \label{eq6}\epsilon_n e^{in\psi_n} =-\frac{\int \int \varepsilon(x,y) r^2 e^{in\phi}dxdy}{\int \int\varepsilon(x,y) r^2 dxdy}\end{equation} 

Here,  $\epsilon_n$'s are parameters characterizing the asymmetry in the distribution, e.g. dipole asymmetry: $\epsilon_1$, participant eccentricity: $\epsilon_2$, participant triangularity,: $\epsilon_3$ etc. $\psi_n$ is the polar angle of the participant plane for n-th order flow. Teaney and Yan \cite{Teaney:2010vd}, from theoretical consideration, argued that for triangular flow, the $r^2$ term in Eq.\ref{eq6} should be replaced by $r^3$,

\begin{equation}  
\epsilon_3 e^{i 3\psi_3}=-\frac{\int \int \varepsilon(x,y) r^3 e^{i 3\phi}dxdy}{\int \int\varepsilon(x,y) r^3 dxdy} \label{eq6a}
\end{equation} 
 
In the present simulation, we have used Eq.\ref{eq6a} to compute triangular flow. Probability distribution of spatial eccentricity ($\epsilon_2$) and triangularity ($\epsilon_3$), in b=8.9 fm Pb+Pb collision   is shown in   Fig.\ref{F1}. Eccentricity distribution  peaks around $\epsilon_2 \approx$ 0.2.
The triangularity parameter $\epsilon_3$ peaks around $\epsilon_3 \approx$ 0.04.
The fluctuations of $\epsilon_3$ however is much larger than that in $\epsilon_2$.
Since elliptic and triangular flows are response respectively to the initial state
eccentricity and   triangularity, larger fluctuations are expected in triangular flow than in elliptic flow.

  \begin{figure}[t]
 \center
 \resizebox{0.30\textwidth}{!}{%
  \includegraphics{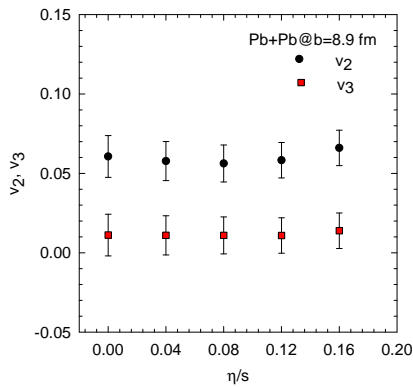} 
}
\caption{(color online) Integrated elliptic and triangular flow  for viscosity to entropy ratio $\eta/s$=0, 0.04, 0.08, 0.12 and 0.16. Event size is $N_{event}$=500.}
\label{F3}
\end{figure}

\section{Viscous effects on elliptic and triangular flow } 

For fluid viscosity to entropy ratio $\eta/s$=0, 0.04, 0.08, 0.12 and 0.16, we have simulated b=8.9 fm Pb+Pb collisions. b=8.9 fm collisions approximately corresponds to 30-40\% collision. In viscous evolution, entropy is generated. To account for the entropy generation, the Gaussian density $\varepsilon_0$ was reduced with increasing viscosity, such that in ideal and viscous fluid, on the average, $\pi^-$ multiplicity remains the same.
In each event, Israel-Stewart's  hydrodynamic equations are solved and from the freeze-out surface, invariant distribution ($\frac{dN}{dyd^2p_T}$) for $\pi^-$ was obtained. 
In analogy to Eq.\ref{eq6}, invariant distribution can be characterised by 'harmonic flow coefficients' \cite{arXiv:1104.0650}.

   \begin{figure}[t]
 \center
 \resizebox{0.30\textwidth}{!}{%
  \includegraphics{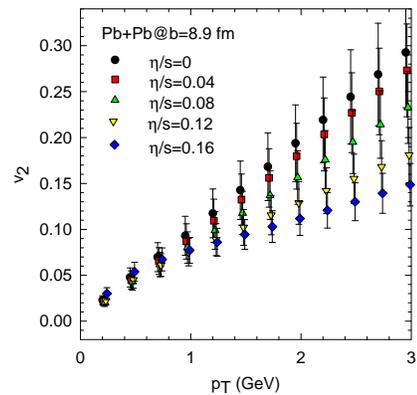} 
}
\caption{(color online) Differential elliptic flow ($v_2$) for viscosity to entropy ratio $\eta/s$=0, 0.04, 0.08, 0.12 and 0.16. Event size is $N_{event}$=60. Data points are shifted marginally to distinguish between them.}
\label{F4}
\end{figure}

\begin{eqnarray}
v_n(y,p_T)e^{in\psi_n(y,p_T)}&=&\frac{\int d\phi e^{in\phi} \frac{dN}{dy  p_Tdp_T d\phi}}  {\frac{dN}{dy p_Tdp_T}}\\
  v_n(y)e^{in\psi_n(y)}&=& \frac{ \int p_T dp_T d\phi e^{in\phi} \frac{dN}{dy p_T dp_T d\phi} } { \frac{dN}{dy} }
\end{eqnarray}

In a boost-invariant version of hydrodynamics, flow coefficients are rapidity independent and in the following, we have dropped the rapidity dependence. 
Present simulations are suitable only for central rapidity, $y\approx$0, where boost-invariance is most justified. 
In the present study, we have used  $N_{event}$=500 events.    
Is sample size $N_{event}$=500 sufficiently large to comment conclusively on fluctuations in $v_2$ and $v_3$?   In Fig.\ref{F2},     simulation results for the  elliptic ($v_2$) and triangular ($v_3$) flow are shown. The filled (open) circles, squares and triangle corresponds to elliptic (triangular) flow for event size $N_{event}$=2500, 500 and 50 respectively. The fluid is assumed to be ideal.  The symbols represent event average and the error bars the variance.
If fluctuations are of statistical origin, one expect the variance in $v_2$ and $v_3$ increase by a factor of $\sim$ 2 between event size 2500 and 500 and by a factor of $\sim$7 between event size 2500 and 50. No such increase is seen in simulations.
As shown in Fig.\ref{F2}
the event averaged elliptic and triangular flow     are
approximately independent of the event size. More importantly, the variances   are
also approximately independent of event size.    The result confirms that fluctuations are systematic rather than statistical. 

We may mention here that the present model of fluctuating initial conditions do not induce large fluctuations in the $\pi^-$  transverse momentum spectra,  $p_T$ spectra remain largely unaffected. The model is also consistent with the experimental observation that in central   collisions, elliptic  and triangular flows are of similar magnitude, but in peripheral collisions, elliptic flow dominates.

In Fig.\ref{F3},  viscosity dependence of integrated elliptic and triangular flow is shown. With initial conditions both $v_2$ and $v_3$  fluctuates. However, fluctuations in $v_2$ is not large, $\sim$20\%. Within the fluctuations, integrated $v_2$ does not show any viscosity dependence. Triangular flow fluctuates more strongly than the elliptic flow. Uncertainty in $v_3$ is $\sim$50\%.  Larger fluctuations in $v_3$ is not unexpected. As shown in Fig.\ref{F1}, triangularity fluctuates more strongly than the eccentricity. It is then expected that the $v_3$ will fluctuate more strongly than $v_2$. Integrated triangular flow also does not show any viscosity dependence.


 \begin{figure}[t]
 \center
 \resizebox{0.30\textwidth}{!}{%
  \includegraphics{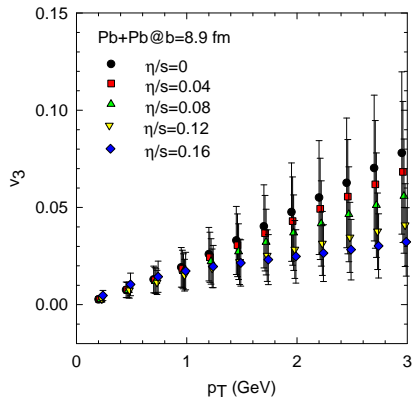} 
}
\caption{(color online) same as in Fig.\ref{F4} but for triangular flow.}
\label{F5}
\end{figure}

In Fig.\ref{F4} and \ref{F5}, differential elliptic and triangular flow is shown as a function of viscosity. Average value of $v_2(p_T)$ and $v_3(p_T)$ decreases with increasing viscosity, however, both elliptic and triangular  flow  fluctuates strongly. In the $p_T$ range 1-3 GeV, fluctuations in $v_2(p_T)$ is $\sim$15-20\%. Fluctuations in $v_3(p_T)$ is even more $\sim$70-80\%.  
The fluctuations in anisotropic flow greatly reduce their efficacy as a diagnostic tool. For example, within the fluctuations, differential elliptic flow does not distinguish between ideal fluid and  fluid with viscosity to entropy ratio 0.08.  Triangular flow is even more insensitive. Fluid viscosity   varying between 0-0.16 is not distinguished. 
 
\begin{figure}[h]
 \center
 \resizebox{0.30\textwidth}{!}{%
  \includegraphics{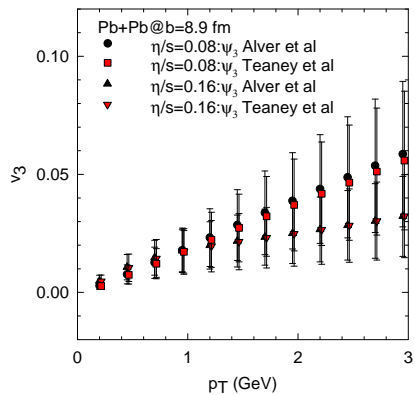} 
}
\caption{(color online) 
Comparison of triangular flow with definition of participant plane due to Teaney and Yan \cite{Teaney:2010vd} and Alver and Rolland \cite{Alver:2010gr}. Simulation results for    fluid viscosity over entropy ratio, $\eta/s$=0.08 and 0.16, are shown. }  \label{F6}
\end{figure}

 Triangular flow depends on the definition of triangularity. In the above simulations, we have used the definition due to Teaney and Yan  \cite{Teaney:2010vd}. As noted earlier,  Alver and Rolland \cite{Alver:2010gr} used an alternate definition for triangularity (see Eq.\ref{eq6}).
It is interesting to compare $v_3$ in the two definitions.  In Fig.\ref{F6}, for  fluid viscosity $\eta/s$=0.08 and 0.16, triangular flow from the two   definitions is compared.   Average triangular flow and its fluctuations are   approximately similar in both the definitions. The result is understood. Triangular flow depend explicitly on the participant plane angle $\psi_3$.    $\psi_3$ is not changed much between the two definition. For example, in the present simulation,
out of 500 events, in $\sim$400 events $\psi_3$ in two definitions differ by less than 10\%.  Marginal change in $\psi_3$ lead to similar triangular flow.

Can the fluctuations in $v_3$ be reduced? As demonstrated earlier, larger event size do not reduce the fluctuations. One of the parameter of the model is the Gaussian width $\sigma$ of the hot spots. Initial density distribution become more asymmetric if $\sigma$ is reduced. We have used $\sigma$=1 fm. Explicit simulations indicate that if $\sigma$ is reduced to  0.5 fm, in the $p_T$ range 1-3 GeV, average of triangular flow  increase by a factor of $\sim$2, the fluctuations however remains largely unaltered. In a more realistic simulation, impact parameter would also fluctuate   and fluctuations in $v_3$ will be even more. Indeed,   if  
the initial state fluctuations are the 'only' source of the triangular flow, large fluctuations in the triangular flow is inevitable. 
Fluctuations will reduce if, apart from initial triangularity, jet quenching, Cerenkov radiation etc. also contribute to the triangular flow. Indeed, hydrodynamical simulations indicate that triangularity in density distribution can be developed due to jet quenching (see Fig.2 of ref.\cite{Chaudhuri:2006qk}), whose response would be the triangular flow.


\section{Conclusions}

In a simple model of fluctuating initial states, we have studied fluctuations in elliptic and triangular flow in ideal and viscous fluid evolution.  It is shown that fluctuations in triangular flow can be very large. Large fluctuation reduces the sensitivity of flow coefficients to viscosity. For example, fluctuations of elliptic flow make it insensitive to   variation of viscosity to entropy ratio in the range $\eta/s$=0-0.08. Fluctuations in triangular flow are even larger and    viscosity to entropy ratio varying between 0-0.16 is not distinguished.  We conclude that if the initial state fluctuations are the only source of triangular flow, triangular flow will be   greatly insensitive to viscosity.


\begin{thebibliography}{99}
\bibitem{Manly:2005zy}
  S.~Manly {\it et al.}  [PHOBOS Collaboration],
  Nucl.\ Phys.\  A {\bf 774}, 523 (2006)
\bibitem{Mishra:2008dm}
  A.~P.~Mishra, R.~K.~Mohapatra, P.~S.~Saumia, A.~M.~Srivastava,
  Phys.\ Rev.\  {\bf C81}, 034903 (2010).
\bibitem{Mishra:2007tw}
  A.~P.~Mishra, R.~K.~Mohapatra, P.~S.~Saumia, A.~M.~Srivastava,
  Phys.\ Rev.\  {\bf C77}, 064902 (2008).
\bibitem{Takahashi:2009na}
  J.~Takahashi, B.~M.~Tavares, W.~L.~Qian, R.~Andrade, F.~Grassi, Y.~Hama, T.~Kodama, N.~Xu,
  Phys.\ Rev.\ Lett.\  {\bf 103}, 242301 (2009).
\bibitem{Alver:2010gr}
  B.~Alver, G.~Roland,
  Phys.\ Rev.\  {\bf C81}, 054905 (2010).
\bibitem{Alver:2010dn}
  B.~H.~Alver, C.~Gombeaud, M.~Luzum, J.~-Y.~Ollitrault,
  Phys.\ Rev.\  {\bf C82}, 034913 (2010).
\bibitem{Teaney:2010vd}
  D.~Teaney, L.~Yan,
  Phys.\ Rev.\  {\bf C83}, 064904 (2011).
\bibitem{:2011vk}
  [ ALICE Collaboration ],
  Phys.\ Rev.\ Lett.\  {\bf 107}, 032301 (2011).


 
 
\bibitem{PHENIXwhitepaper} 
PHENIX Collaboration, K.~Adcox {\it et al.}, 
Nucl. Phys. A {\bf 757} 184 (2005).  
  
\bibitem{STARwhitepaper} 
STAR Collaboration, J. Adams {\it et al.}, 
Nucl. Phys. A {\bf 757} 102 (2005).

\bibitem{Aamodt:2010pa}
  K.~Aamodt {\it et al.}  [The ALICE Collaboration],
  arXiv:1011.3914 [nucl-ex].


\bibitem{Luzum:2008cw}
  M.~Luzum and P.~Romatschke,
  Phys.\ Rev.\  C {\bf 78}, 034915 (2008).


\bibitem{Song:2008hj}
  H.~Song and U.~W.~Heinz,
  J.\ Phys.\ G {\bf 36}, 064033 (2009).
  
\bibitem{Chaudhuri:2009uk}
  A.~K.~Chaudhuri,
  Phys.\ Lett.\  B {\bf 681}, 418 (2009).


\bibitem{Chaudhuri:2009hj}
  A.~K.~Chaudhuri,
  J.\ Phys.\ G {\bf G37}, 075011 (2010).

 

\bibitem{Roy:2011xt}
  V.~Roy and A.~K.~Chaudhuri, Phys. Lett. B (in press)
  arXiv:1103.2870 [nucl-th].
\bibitem{Schenke:2011tv}
  B.~Schenke, S.~Jeon, C.~Gale,
  Phys.\ Lett.\  {\bf B702}, 59-63 (2011).
\bibitem{Bozek:2011wa}
  P.~Bozek,
  Phys.\ Lett.\  {\bf B699}, 283-286 (2011).
\bibitem{Song:2011qa}
  H.~Song, S.~A.~Bass, U.~Heinz,
  Phys.\ Rev.\  {\bf C83}, 054912 (2011).

  

  
\bibitem{Song:2008si}
  H.~Song and U.~W.~Heinz,
  Phys.\ Rev.\  C {\bf 78}, 024902 (2008).

  
\bibitem{Chaudhuri:2008sj} A.~K.~Chaudhuri,
 arXiv:0801.3180 [nucl-th].
 
 
\bibitem{Aoki:2006we}
  Y.~Aoki, G.~Endrodi, Z.~Fodor, S.~D.~Katz and K.~K.~Szabo,
  Nature {\bf 443}, 675 (2006)

\bibitem{Aoki:2009sc}
  Y.~Aoki, S.~Borsanyi, S.~Durr, Z.~Fodor, S.~D.~Katz, S.~Krieg and K.~K.~Szabo,
  JHEP {\bf 0906}, 088 (2009)
\bibitem{Borsanyi:2010cj}
  S.~Borsanyi {\it et al.},
  JHEP {\bf 1011}, 077 (2010)
  [arXiv:1007.2580 [hep-lat]].
 
 \bibitem{Fodor:2010zz}
  Z.~Fodor,
  J.\ Phys.\ Conf.\ Ser.\  {\bf 230} (2010) 012013.
 

\bibitem{CasalderreySolana:2009uk}
  J.~Casalderrey-Solana, U.~A.~Wiedemann,
  Phys.\ Rev.\ Lett.\  {\bf 104}, 102301 (2010).
  [arXiv:0911.4400 [hep-ph]].



\bibitem{Bozek:2009dt}
  P.~Bozek,
  Acta Phys.\ Polon.\  {\bf B41}, 837 (2010).



   
\bibitem{Chaudhuri:2009yp}
  A.~K.~Chaudhuri,
  Phys.\ Lett.\  B {\bf 692}, 15 (2010)

\bibitem{Bhalerao:2011bp}
  R.~S.~Bhalerao, M.~Luzum, J.~-Y.~Ollitrault,
 [arXiv:1107.5485 [nucl-th]].

\bibitem{Chaudhuri:2006qk}
  A.~K.~Chaudhuri,
  Phys.\ Rev.\  C {\bf 75}, 057902 (2007).
  
\bibitem{arXiv:1104.0650} 
  Z.~Qiu and U.~W.~Heinz,
  Phys.\ Rev.\ C\ {\bf 84}, 024911  (2011)
  [arXiv:1104.0650 [nucl-th]].


\end{thebibliography}
\end{document}